\documentclass[aps,prb,twocolumn,amsmath,showpacs,endfloats]{revtex4}
\usepackage{graphicx}

\begin{document}

\title{Local and average fields inside surface-disordered 
waveguides: Resonances in the one-dimensional Anderson 
localization regime}

\author{Jos\'e A. S{\'a}nchez-Gil}
\email{j.sanchez@iem.cfmac.csic.es}
\affiliation{Instituto de Estructura de la Materia,
Consejo Superior de Investigaciones Cient{\'\i}ficas,
Serrano 121, E-28006 Madrid, Spain
}
\author{Valentin Freilikher}
\affiliation{The Jack and Pearl Resnick Institute of Advanced Technology, 
Department of Physics, Bar-Ilan University, Ramat-Gan 52900, Israel
}
\date{\today }

\begin{abstract}

We investigate the one-dimensional propagation of waves in the
Anderson localization regime, for a single-mode, surface disordered
waveguide. We make use of both an analytical formulation and rigorous
numerical simulation calculations. The occurrence of anomalously large
transmission coefficients for given realizations and/or frequencies is
studied, revealing huge field intensity concentration inside the
disordered waveguide.  The analytically predicted $s$-like dependence
of the average intensity, being in good agreement with the numerical
results for moderately long systems, fails to explain the intensity
distribution observed deep in the localized regime. The average
contribution to the field intensity from the resonances that are above
a threshold transmission coefficient $T_{c}$ is a broad distribution
with a large maximum at/near mid-waveguide, depending universally (for
given $T_{c}$) on the ratio of the length of the disorder segment to
the localization length, $L/\xi$. The same universality is observed in
the spatial distribution of the intensity inside typical (non-resonant
with respect to the transmission coefficient) realizations, presenting
a s-like shape similar to that of the total average intensity for
$T_{c}$ close to 1, which decays faster the lower is $T_{c}$.
Evidence is given of the self-averaging nature of the random quantity
$\log[I(x)]/x\simeq -1/\xi$.  Higher-order moments of the intensity
are also shown.

\end{abstract}

\maketitle

\section{INTRODUCTION}

\label{sec_int}

There are two well-known manifestations of strong localization of
classical waves in one-dimensional (1D) open disordered systems:
exponentially small (with respect to the length of the system)
transmission through typical (most probable) random realizations, and
high transparency at rare (exponentially low-probable) ones. The high
transparency is due to the so called stochastic resonances that are
accompanied by large concentration (localization) of energy in
relatively small areas inside the system. It was shown in Ref.
\onlinecite{frisch} that in a semi-infinite random medium the wave
amplitude at the resonances can exceed (with nonzero probability) any
given value. In the 80s this phenomenon had been studied intensively
as applied to electrons, light, elastic, and acoustical waves
\cite{papanico,Azbel} (see also Refs.
\onlinecite{klbook,lifsh,sheng,frei} and references therein).  In the
last few years, after a long hiatus, interest in stochastic resonances
in random media has rekindled in the context of random lasing
\cite{letho,cao,natu,lag01}, wherein resonances might play the role of
effective confining cavities inducing lasing action when gain is
introduced.

We investigate the one-dimensional propagation of electromagnetic (EM)
waves in the strong localization regime. In particular, the occurrence
of anomalously large transmission coefficients for given realizations
and/or frequencies (resonant or quasi-transparent realizations) is
studied, with emphasis on the field intensity distributions along the
direction of propagation. For that purpose, we make use of both an
analytical formulation and rigorous numerical simulation calculations.

We consider a single-mode waveguide with randomly rough walls. This
structure, being a typical example, of a one-dimensional disordered
system, has the advantage that it can be easily prepared using
standard equipment (microwave waveguides or fiber optics), and enables
(unlike a random stack of dielectric layers) to directly measure the
wave field inside the structure. Similar multi-mode systems have been
studied in recent years to investigate various localization and
transport phenomena appearing in the propagation of waves through
disordered media
\cite{prl98,mnvprl98,prb99,mnvprl00,mnvprl01,mnvprl02,izma02}.

Our numerical calculations exploit the invariant embedding equation
formulation for a multi-mode surface-disordered waveguide
\cite{prl98,prb99,emb}, which we have extended to account for the
field inside the disordered region. The numerical results are compared
with analytical formulas obtained by using the invariant embedding
method and averaging over rapid phase variations
\cite{emb,klbook}. Both methods are described in
Sec.~\ref{sec_eq}. Local and average field intensities are presented
Secs.~\ref{sec_reso} and~\ref{sec_mean}, respectively; the conclusions
drawn from them are summarized in Sec.~\ref{sec_con}.

\section{SCATTERING MODEL}

\label{sec_eq}

\subsection{Field distribution outside the disordered region: 
Reflection and transmission amplitudes}

\begin{figure}[h]
\includegraphics[width=.9\columnwidth]{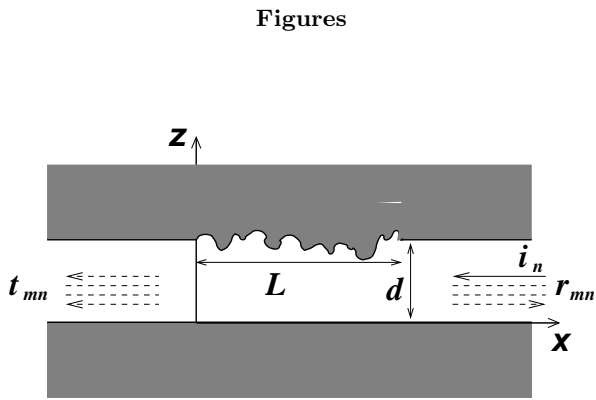}
\caption{Illustration of the scattering geometry of the 
surface-disordered waveguide. }
\label{fig_sca}
\end{figure}

The scattering geometry is depicted in Fig.~\ref{fig_sca}. We seek for
solutions to the scalar Helmholtz equation %\begin{equation}
%\left( \Delta +k^2\right) \Psi \left( {\bf R}\right) =0,
%\label{eq_we}
%\end{equation}
in the form (outside the region $0\leq x\leq L$): 
\begin{subequations}
\begin{eqnarray}
\Psi _n\left( x,\mathbf{r}\right) = {\displaystyle \sum_m
k^{-1/2}_m\chi_m\left( \mathbf{r}\right) e^{-ik_mx}t_{mn},} \quad x<0, && \\
\Psi _n\left( x,\mathbf{r}\right) = {\displaystyle \Psi^0 _n\left( x,\mathbf{%
r}\right)} \hspace*{4cm} &&  \notag \\
{\displaystyle +\sum_mk^{-1/2}_m\chi_m\left( \mathbf{r}\right)
e^{ik_mx}r_{mn}, } \quad x>L, &&
\end{eqnarray}
with 
\begin{eqnarray}
\Psi^0 _n\left( x,\mathbf{r}\right)=k^{-1/2}_n\chi _n(\mathbf{r}) e^{-ik_nx}.
\end{eqnarray}\label{eq_inc}\end{subequations}

The indexes ``$m,n$'' correspond to the outgoing and incoming modes,
respectively. $\chi _n\left( \mathbf{r}\right)$ are the eigenfunctions of
the transverse wave equation, characterized by transverse momentum 
{\boldmath $\kappa_n$}, so that the longitudinal wavevector component $k_n$
(along the propagation direction) is 
\begin{equation}
k_n=\left[(\omega/c)^2-\kappa _n^2\right]^{1/2},
\end{equation}
with $\omega$ being the wave frequency.

We consider the Dirichlet boundary condition on a slightly perturbed
waveguide surface, {\boldmath $\zeta$} denoting the random perturbation, and
expand it about the unperturbed surface $\mathbf{R}=\mathbf{R_s}$, which is
translationally invariant along the $x$-axis [$\mathbf{R}=\left( x,\mathbf{r}%
\right)$], so that: 
\begin{subequations}
\begin{eqnarray}
\Psi\left(\mathbf{R}=\mathbf{R_s}\right)= 0, \text{ for } x<0 \text{ and }
x>L, \hskip 0.5 in && \\
{\displaystyle =-\mbox{\boldmath $\zeta$}\left(\mathbf{R}\right)\cdot \frac{%
\partial \Psi \left(\mathbf{R} \right) }{\partial \mathbf{R}},} \text{ for }
0\leq x\leq L. & &
\end{eqnarray}\label{eq_bc}\end{subequations}
Alternatively, the latter boundary condition can be associated to a
waveguide surface with a random admittance.

It can be shown that the matrices of reflection and transmission
coefficients satisfy the following differential equations \cite{prb99}: 
\begin{subequations}
\label{eq_iee}
\begin{eqnarray}
&& \frac{d\widehat{r}}{dL}=\frac i2\left( e^{-i\widehat{k}L}+\widehat{r} e^{i%
\widehat{k}L}\right) \widehat{v}\left( e^{-i\widehat{k}L}+ e^{i\widehat{k}L}%
\widehat{r}\right) ,  \label{eq_ier} \\
&& \frac{d\widehat{t}}{dL}=\frac i2\widehat{t}e^{i\widehat{k}L}\widehat{v}
\left( e^{-i\widehat{k}L}+e^{i\widehat{k}L}\widehat{r}\right) ,
\label{eq_iet}
\end{eqnarray}
with $\widehat{k}=diag\left( k_n\right) $ and 
\end{subequations}
\begin{eqnarray}
v_{mn}=\oint ds\,\phi _m\left( s\right) \zeta \left( L,s\right) \phi
_n\left(s\right) ,  \notag \\
\phi _n\left( s\right) =k^{-1/2}_n\mathbf{n}(\mathbf{r_s})\cdot \left[ \frac{%
\partial \chi _n\left( \mathbf{r}\right) }{\partial \mathbf{r}}\right] _{%
\mathbf{r}=\mathbf{r_s}};  \notag  \label{eq_vmn}
\end{eqnarray}
it has been assumed that {\boldmath $\zeta$}$=\zeta\mathbf{n}$. The explicit
form of the differential $ds$ over the cross section (oriented) surface
element $d\mathbf{s}=\mathbf{n}ds$ depends on the geometry under
consideration. The reflection and transmission intensities are defined by: 
\begin{eqnarray}
R_{mn}=\left| r_{mn}\right| ^2, T_{mn}=\left| t_{mn}\right| ^2,
\end{eqnarray}
which yield the intensity coupled into the $mth$ outgoing channel in
reflection and transmission, respectively, for a given $nth$ incoming
channel.

\subsection{Field distribution inside the disordered region}

By invoking Green's theorem, the expression for the field inside the
waveguide ($0\leq x\leq L$) can be written as: 
\begin{eqnarray}
&&\Psi _{n}\left( x,\mathbf{r}\right) =\Psi _{n}^{0}\left( x,\mathbf{r}
\right)  \notag \\
&&+\int\limits_{0}^{L}dx^{\prime }\oint ds^{\prime }\Psi _{n}\left(
x^{\prime },\mathbf{r_{s}^{\prime }}\right) \frac{\partial G_{0}\left(
x^{\prime },x;\mathbf{r_{s}^{\prime }},\mathbf{r}\right) }{\partial
n^{\prime }},  \label{eq_gti2}
\end{eqnarray}
where $\Psi _{n}^{0}=k_{n}^{-1/2}\chi _{n}(\mathbf{r})e^{-ik_{n}x}$.
Substituting the Green's function 
\begin{equation}
G_{0}\left( x,\mathbf{r};x^{\prime }\mathbf{r^{\prime }}\right)
=\sum_{m=1}^{N}(2ik_{m})^{-1}\chi _{m}\left( \mathbf{r}\right) \chi
_{m}\left( \mathbf{r^{\prime }}\right) e^{ik_{m}\left| x-x^{\prime }\right| }
\end{equation}
into Eq.~(\ref{eq_gti2}), we end up with the following expression for the
scattered field inside: 
\begin{eqnarray}
\Psi _{n}^{sc}\left( x,\mathbf{r}\right) =\Psi _{n}\left( x,\mathbf{r}
\right) -\Psi _{n}^{0}\left( x,\mathbf{r}\right) \hskip1in &&  \notag \\
=\int\limits_{0}^{L}dx^{\prime }\oint ds^{\prime }\Psi _{n}\left( x^{\prime
},\mathbf{r_{s}^{\prime }}\right) \hskip.8in &&  \notag \\
\times \sum_{m=1}^{N}(2ik_{m})^{-1}\chi _{m}(\mathbf{r})\frac{\partial \chi
_{m}(\mathbf{r_{s}^{\prime }})}{\partial n^{\prime }}e^{ik_{m}|x-x^{\prime
}|}, &&
\end{eqnarray}
where $\Psi _{n}^{0}=k_{n}^{-1/2}\chi _{n}(\mathbf{r})e^{-ik_{n}x}$.
Rearranging the integrand, and handling the phase factor $%
e^{ik_{m}|x-x^{\prime }|}$ appropriately by splitting the integral along the
waveguide $\int\limits_{0}^{L}=\int\limits_{0}^{x}+\int\limits_{x}^{L}$, one
obtains: 
\begin{eqnarray}
\Psi _{n}^{sc}\left( x,\mathbf{r}\right) =\sum_{m=1}^{N}(2ik_{m})^{-1/2}\chi
_{m}(\mathbf{r})\hskip1.1in &&  \notag \\
\times \left\{ \int\limits_{0}^{x}dx^{\prime }e^{ik_{m}(x-x^{\prime })}\oint
ds^{\prime }\Psi _{n}\left( x^{\prime },\mathbf{r_{s}^{\prime }}\right) 
\frac{\partial \chi _{m}(\mathbf{r_{s}^{\prime }})}{\partial n^{\prime }}
\right. &&  \notag \\
\left. +\int\limits_{x}^{L}dx^{\prime }e^{-ik_{m}(x-x^{\prime })}\oint
ds^{\prime }\Psi _{n}\left( x^{\prime },\mathbf{r_{s}^{\prime }}\right) 
\frac{\partial \chi _{m}(\mathbf{r_{s}^{\prime }})}{\partial n^{\prime }}
\right\} . &&
\end{eqnarray}
Factoring out the phase factors defining waves propagating right and left,
and splitting again the integral of the second term, we get 
\begin{eqnarray}
\Psi _{n}^{sc}\left( x,\mathbf{r}\right) =\sum_{m=1}^{N}\frac{\chi _{m}
(\mathbf{r})}{k_{m}^{1/2}}\left\{ \frac{e^{ik_{m}x}}{2ik_{m}^{1/2}}\right.
\hskip1.2in &&  \notag \\
\times \int\limits_{0}^{x}dx^{\prime }\oint ds^{\prime }\Psi _{n}\left(
x^{\prime },\mathbf{r_{s}^{\prime }}\right) \frac{\partial \chi _{m}
(\mathbf{r_{s}^{\prime }})}{\partial n^{\prime }}e^{-ik_{m}x^{\prime }}
+\frac{e^{-ik_{m}x}}{2ik_{m}^{1/2}} &&  \notag \\
\left. \left( \int\limits_{0}^{L}dx^{\prime }-\int\limits_{0}^{x}dx^{\prime
}\right) \oint ds^{\prime }\Psi _{n}\left( x^{\prime },\mathbf{r_{s}^{\prime
}}\right) \frac{\partial \chi _{m}(\mathbf{r_{s}^{\prime }})}{\partial
n^{\prime }}e^{ik_{m}x^{\prime }}\right\}. &&
\end{eqnarray}
At this point, we define the local amplitudes of the
scattered waves propagating along the $x$ axis in positive and
negative directions, respectively, $\rho (x)$ and $\tau (x)$:
\begin{subequations}
\begin{eqnarray}
\rho _{mn}^{L}(x)=(2ik_{m}^{1/2})^{-1}\hskip1.5in &&  \notag \\
\times \int\limits_{0}^{x}dx^{\prime }\oint ds^{\prime }\Psi _{n}
\left(x^{\prime },\mathbf{r_{s}^{\prime }}\right) \frac{\partial 
\chi _{m}(\mathbf{r_{s}^{\prime }})}{\partial n^{\prime }}
e^{-ik_{m}x^{\prime }}, && \\
\tau _{mn}^{L}(x)=t_{mn}(L)-\delta _{mn}-(2ik_{m}^{1/2})^{-1}
\hskip.5in && \notag \\
\times \int\limits_{0}^{x}dx^{\prime }\oint ds^{\prime }\Psi _{n}
\left(x^{\prime },\mathbf{r_{s}^{\prime }}\right) \frac{\partial 
\chi _{m}(\mathbf{r_{s}^{\prime }})}{\partial n^{\prime }}
e^{ik_{m}x^{\prime }}, &&
\end{eqnarray}\label{eq_rtin}\end{subequations}
so that 
\begin{eqnarray}
\Psi^{sc}_n\left( {\bf R}\right)= \sum_{m=1}^{N}
 k_m^{-1/2}\chi_n({\bf r}) \hskip 1.2in  &&\nonumber \\ 
 \times\{ \rho^L_{mn}(x)e^{ik_mx}+\tau^L_{mn}(x)e^{-ik_m x}\}.
\label{eq_scin}\end{eqnarray}

Then, by differentiating Eqs.~(\ref{eq_rtin}), and taking into account the
boundary condition~(\ref{eq_bc}) in the integrands, with the aid of Eq.~
(\ref{eq_scin}) again, a set of coupled differential equations for $\rho
_{mn}^{L}(x)$ and $\tau _{mn}^{L}(x)$ is derived: 
\begin{subequations}
\begin{eqnarray}
\frac{d\widehat{\rho }^{L}}{dx} &=&\frac{i}{2}e^{-i\widehat{k}x}
\widehat{v}\left[ e^{-i\widehat{k}x}(\widehat{I}+\widehat{\tau }^{L})
+e^{i\widehat{k}x}\widehat{\rho }^{L}\right], \\
\frac{d\widehat{\tau }^{L}}{dx} &=&-\frac{i}{2}e^{i\widehat{k}x}
\widehat{v}\left[ e^{-i\widehat{k}x}(\widehat{I}+\widehat{\tau }^{L})
+e^{i\widehat{k}x}\widehat{\rho }^{L}\right].
\end{eqnarray}\label{eq_ode_rtin}\end{subequations}
The corresponding boundary conditions satisfied by $\rho _{mn}^{L}(x)$ and
$\tau _{mn}^{L}(x)$ at the end points of the waveguide are: 
\begin{eqnarray}
\rho _{mn}^{L}(x &=&0)=0,\tau _{mn}^{L}(x=0)=t_{mn}(L)-\delta _{mn}; \\
\rho _{mn}^{L}(x &=&L)=r_{mn}(L),\tau _{mn}^{L}(x=L)=0.  \label{eq_bcin}
\end{eqnarray}

\subsection{Numerical calculations}

We have chosen for the numerical simulations the same geometry as in
Ref.  \onlinecite{prb99}: two parallel, perfectly reflecting planes at
$z=0$ and $z=d$ with random deviations $z=\zeta \left( x\right) $
given by a 1D stochastic process with Gaussian statistics
with zero mean and a Gaussian surface power spectrum
\begin{equation}
g(Q)=\delta ^{2}\pi ^{\frac{1}{2}}a\exp \left( -(Qa)^{2}/4\right) ,
\label{eq_ps}
\end{equation}
where $\delta $ is the RMS height and $a$ is the transverse correlation
length. The corresponding transverse eigenfunctions are thus given by
\begin{equation}
\chi _{n}\left( z\right) =(2/d)^{1/2}\sin \left( \kappa _{n}z\right)
,\;\kappa _{n}=\pi n/d,
\end{equation}
and the impurity matrix~(\ref{eq_vmn}) by
\begin{equation}
v_{mn}\left( L\right) =\frac{2}{d}\frac{\kappa _{n}\kappa _{m}}
{(k_{n}k_{m})^{1/2}}\zeta \left( L\right) .  \label{eq_im}
\end{equation}
In order to model 1D wave propagation in our calculations, the
waveguide supports only one mode, its thickness $d$ being such that
$\bar{\omega} \equiv \omega d/(2\pi c)\approx 0.75$. Consequently, all
subscripts referring to mode indexes are suppressed hereafter.

The linear differential equations for the reflection and transmission
amplitudes (\ref{eq_iee}) are solved numerically by means of the
Runge-Kutta method; this is done for a given realization
$\zeta (x)$ from $L=0$ up to a maximum length $L=L_{max}$
(cf. Ref.~\onlinecite{prb99}). Then, for a fixed length of the
disordered segment $L$, the same standard numerical techniques are
employed for the system of first-order differential
Eqs.~(\ref{eq_ode_rtin}) in order to obtain the \textit{local}
reflection and transmission amplitudes, with the help of the boundary
conditions~(\ref{eq_bcin}) involving the reflection and transmission
amplitudes [$r(L),t(L)$], previously obtained. Finally, the field
intensity is calculated from the incident and scattered fields inside
[Eqs.~(\ref{eq_inc}c) and~(\ref{eq_scin})]:
\begin{equation}
I(x)=\mid \Psi (x,d/2)/\Psi ^{0}(x,d/2)\mid ^{2}.  \label{eq_ix}
\end{equation}

\subsection{Analytical approach: Rapid phase averaging }

To calculate analytically the average intensity, $\left\langle
I(x)\right\rangle $, inside a one-dimensional disordered system we
introduce the function
\begin{equation}
R(x)=\frac{\rho (x)}{\tau (x)+1}\exp (2ikx), \label{kl-fr}
\end{equation}
that satisfies the nonlinear equation 
\begin{equation}
i\frac{dR(x)}{dx}=-2kR(x)+\frac{V(x)}{2k}[1+R(x)]^{2},\text{ \ }
\label{anal2}
\end{equation}
where $k$ is the longitudinal wave number of the propagating mode 
($k=\left[(\omega /c)^{2}-\left( \pi/d\right) ^{2}\right] ^{1/2})$.
The random scattering potential, $V(x)$, in the case under
consideration, i.e.  in a single-mode waveguide with a randomly rough
surface, has the form:
\begin{equation}
V(x)=-\frac{2\pi }{d^{3}}\zeta \left( x\right) .  \label{potent}
\end{equation}
Then the intensity can be expressed as \cite{klbook}
\begin{equation}
I(x)=I_{0}\left[ 1-\left| R(x)\right| ^{2}\right] \frac{1+\left| 
R(x)\right|^{2}+2\Re(R(x))}{1-\left| R(x)\right| ^{2}}. \label{anal1}
\end{equation}

Obviously, $R(L)=r(L)\exp (2ikL)$ where $r(L)$ is the total reflection
coefficient of a single-mode waveguide defined by Eq. (\ref{eq_ier}).
It is convenient, following Ref. \onlinecite{klbook}, to introduce two
functions, $u(x)$ and $\varphi (x)$, so that
\begin{equation}
R(x)=\sqrt{\frac{u(x)-1}{u(x)+1}}\exp \left[ i\varphi (x)\right] .
\label{anal3}
\end{equation}
Substitution of Eq. (\ref{anal3}) into Eq. (\ref{anal1}) yields
\begin{equation}
I(x)=\frac{2I_{0}}{u(L)+1}\left[ u(x)+\sqrt{u^{2}(x)-1}\cos \varphi (x)
\right] . \label{anal5}
\end{equation}

If the scattering is weak enough, so that $l_{scat}\gg\lambda$, the
random phase, $\varphi (x)$, is uniformly distributed over $[0,2\pi]$;
We have verified this assumption through numerical calculations of the
probability density function of $\varphi (x)$ (not shown here), which
indeed yield a uniform distribution in all cases studied below.
Obviously, to get rid of the rapid (on the scale of order of $\lambda
)$ oscillations of the phase one has to integrate (average)
Eq. (\ref{anal5}) over an interval $\pounds$ that satisfies the
inequality $\lambda\ll\pounds\ll l_{scat}.$ This rapid phase averaging
(RPA) yields
\begin{equation}
I(x)=\frac{2I_{0}u(x)}{u(L)+1}.  \label{anal7}
\end{equation}

The two-point probability distribution function,
$p_{2}(u_{L},L;u_{x},x)$, necessary for the ensemble averaging of the
intensity $I(x)$, Eq. (\ref{anal7}), can be also calculated under the
assumptions that $l_{scat}\gg\lambda$ and that the scattering
potential is a $\delta$-correlated Gaussian random process such that
\begin{equation}
\langle \delta V(x)\delta V(x^{^{\prime }})\rangle =D
    \delta (x-x^{^{\prime}}). \label{anal6}
\end{equation}
Then, the smoothed (RPA) mean intensity distribution inside a
one-dimensional random system can be presented in the form
\cite{klbook}
\begin{eqnarray}
\left\langle I(x)\right\rangle =\pi \exp [D(x-L/4)]\int_{0}^{\infty }d\mu 
\frac{\sinh \mu \pi }{\cosh ^{2}\mu \pi } &&  \notag \\
\times \exp (-\mu ^{2}DL)\left( \cos 2\mu Dx+\frac{\sin 2\mu Dx}{2\mu }
\right). &&  \label{eq_ai_rpa}
\end{eqnarray}

In what follows, comparisons with the numerical calculations will be
made on the basis of the average (macroscopic) properties, regardless
of the (microscopic) details of the disorder. Namely, the localization
length $\xi$, defined from $\langle \log T\rangle \sim -L/\xi $, will
be used as matching parameter, which in this RPA approach is given by
$\xi=D^{-1}$.

\section{SINGLE REALIZATIONS: RESONANCES}
\label{sec_reso}

\begin{figure}[h]
\includegraphics[width=\columnwidth]{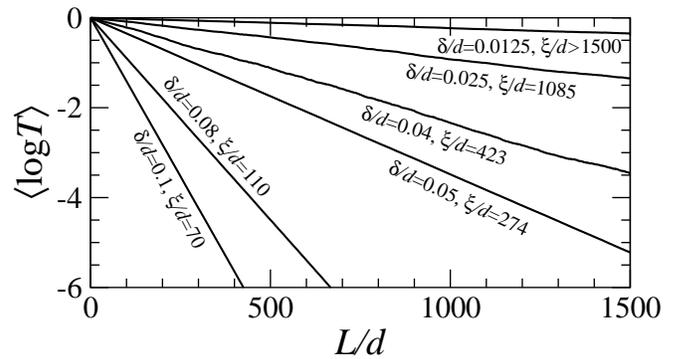}
\caption{Average logarithm of the transmission coefficient over
$N=10^5$ realizations as a function of the disorder length $L$ for
surface roughness parameters $a/d=0.2$ and
$\delta/d=0.0125,0.025,0.04,0.05,0.08,0.1$.  The localization lengths
$\xi$ resulting from fits to linear decays are shown. }
\label{fig_alnt}
\end{figure}

First, we identify the roughness and waveguide parameters that lead to
the onset of Anderson localization. This is done in
Fig.~\ref{fig_alnt} by plotting the length dependence of $\langle \log
T\rangle $ at frequency $\bar{\omega}=0.75$ (single mode) for several
RMS heights $\delta$ and fixed correlation length $a/d=0.2$. The
resulting linear decay is the fingerprint of Anderson localization,
the decay rate yielding the localization length.
The fitted values of $\xi $
for each $\delta $ are included in Fig.~\ref{fig_alnt}.

\begin{figure}[h]
\includegraphics[width=\columnwidth]{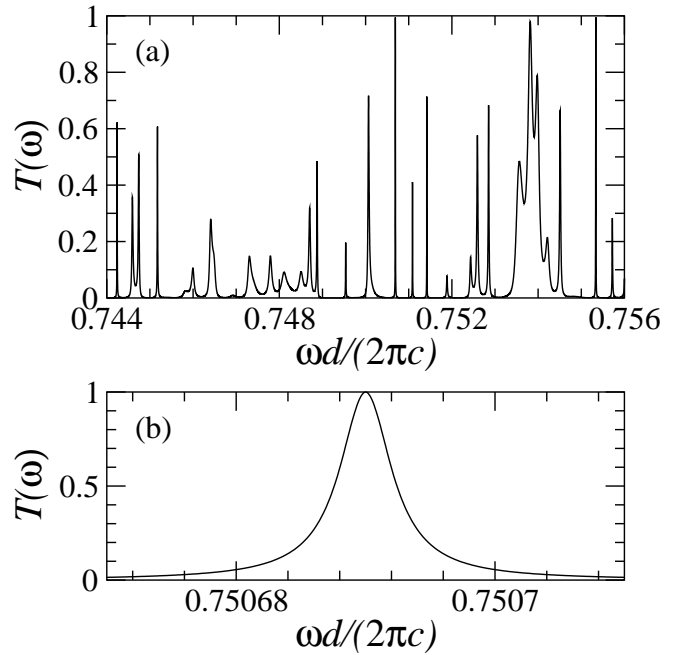}
\caption{(a) Spectral dependence of the transmission coefficient for a
given disorder realization with $a/d=0.2$, $\delta/d=0.05$, and
$L/d=1500$ in a narrow frequency range showing several resonant
frequencies. (b) A single resonance is zoomed in and fitted to a
Lorentzian (dashed curve, indistinguishable from the numerical
result. }
\label{fig_tw}
\end{figure}

With the aid of the latter results, we choose a set of parameters that
ensure the 1D Anderson localization regime: $a/d=0.2$, $\delta
/d=0.05$, and $L=1500d\approx 5.5\xi $. We then calculate the
frequency dependence (in a narrow frequency range) of the transmission
coefficient $T(\omega )$ for a given realization, as shown in
Fig.~\ref{fig_tw}(a). Extremely large fluctuations are observed with
narrow spikes appearing over a fairly negligible background. The
latter background yields the expected response at typical (high
probability) frequencies, since $\langle \log T(\bar{\omega}
=0.75)\rangle \sim -5.2$ and $\langle T(\bar{\omega}=0.75)\rangle \sim
0.062$. The low-probability peaks in Fig.~\ref{fig_tw}(a) correspond
to narrow \textit{resonances or quasi-transparent frequencies} at
which the transmission coefficient can be even 1.

The transmission in the vicinity of one such transparent frequency
($\bar{\omega}_0=0.75069$) is presented in detail in
Fig.~\ref{fig_tw}(b). Note the frequency scale, revealing how narrow
the resonance is. By fitting the numerical result to a Lorentzian
[also shown in Fig.~\ref{fig_tw}(b)], we obtain the half-width at
half-maximum $\Gamma/\omega_0 \approx 2.4\times 10^{-6}$. Resonances
behave like high-finesse cavity modes with large associated
$Q$-factors ($\sim 3\times 10^5$), which may lead to practical
applications as in random lasing \cite{cao,natu,lag01}.

\begin{figure}[h]
\includegraphics[width=\columnwidth]{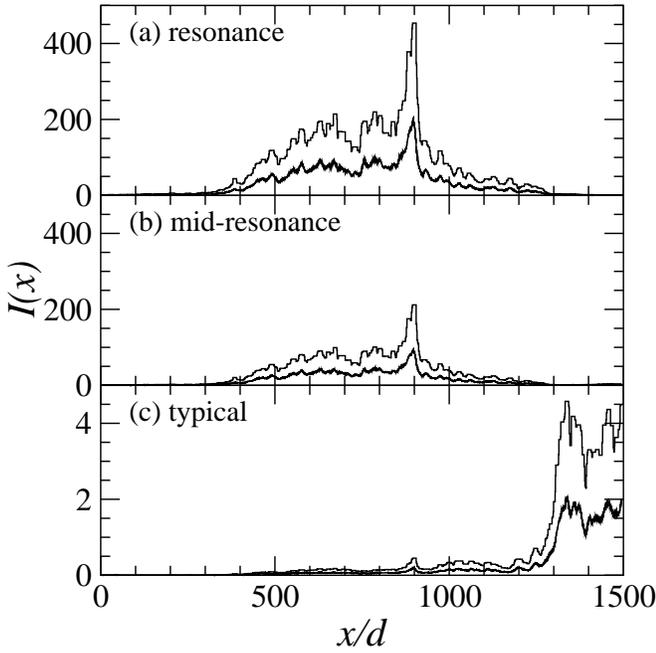}
\caption{Field intensity along the disorder realization used in
Fig.~{\ref{fig_tw}}(b) (with $a/d=0.2$, $\delta/d=0.05$, and
$L/d=1500$) at (a) $\bar{\omega}\equiv \omega d/(2\pi c)=$0.75069
(resonance), (b) $\bar{\omega}=$0.7506875 (mid-resonance), and (c)
$\bar{\omega}=$0.7506 (out of resonance, typical). To suppress rapid
spatial oscillations, the envelopes (upper curves) and spatial
averages (lower curves) are shown. The incident wave is coming from
the right end. }
\label{fig_ix}
\end{figure}

The field intensities inside the waveguide for frequencies at the
resonance, mid-resonance, and out-of-resonance [$\bar{\omega}=0.75069,
0.7506875$, and 0.7506, respectively, in Fig.~\ref{fig_tw}(b)] are
shown in Fig.~ \ref{fig_ix}, where the envelope and average of $I(x)$
over rapid oscillations (period $\sim\pi/k$) are plotted. The incident
mode impinges on the disordered segment at $x=L$ propagating from
right (positive $x$ axis) to left (negative $x$ axis). At resonance
[see Fig.~\ref{fig_ix}(a)], high intensity concentration takes place
over a  region around the center of the disordered segment of the
waveguide ($I\sim 200$ with a peak of $I\sim 400$), its particular 
shape being a characteristic feature of the given resonance. The field
intensity at the end points (not discernible in the figure) is
$I(x=0,L)=1$, as expected ($T=1, R=0$). At mid-resonance [see
Fig.~\ref{fig_ix}(b)], the field intensity distribution maintains its
shape, but the overall height is decreased by nearly a factor of
2. The reflected and transmitted coefficients are retrieved at the end
points: $I(x=L)=\mid 1+r\exp(ikL)\mid^2$ (envelope $\approx 2.25$ and
mean $\approx 1.5$) and $I(x=0)=T\approx 0.25$. 

In contrast, an absolutely different behavior has been observed away
from resonance, i.e. at typical (non-transparent) frequencies (or
realizations), as seen in Fig.~\ref{fig_ix}(c). The field energy is not
localized, but decays from its initial value $I(x=L)$
(envelope $\approx 4$ and mean $\approx 2$) to the exponentially small
value $I(x=0)=T\sim \exp(-L/\xi)$. 

\section{TOTAL, TYPICAL AND RESONANT AVERAGE FIELDS}

\label{sec_mean}

We now turn to the analysis of the ensemble average of the field
intensity $\langle I(x)\rangle$ along the disordered region. Numerical
simulation calculations are carried out for fixed $\bar{\omega}=0.75$,
$L$, and statistical parameters of the roughness. Averages have been
done over $N=10^5 $ realizations, separating typical and resonant
realizations according to a threshold value of the mean transmission
coefficient $T_c$.

\begin{figure}[h]
\includegraphics[width=\columnwidth]{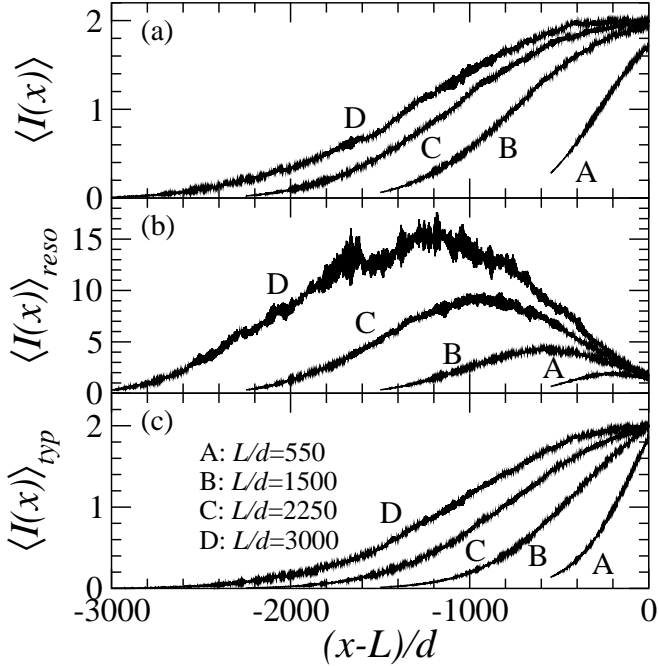}
\caption{(a) Spatial distributions of average field intensity
($N=10^5$ realizations) for $a/d=0.2$, $\delta/d=0.05$, and disorder
lengths: $L/d=$550 (curves A), 1500 (curves B), 2250 (curves C), and
3000 (curves D).  All curves have been shifted to make coincide the
incoming ends at $x=0$.  The contributions from resonant realizations
(with $T\geq T_c=0.1$) and the remaining typical realizations are
shown in (b) and (c), respectively. }
\label{fig_aix}
\end{figure}

Figure~\ref{fig_aix}(a) shows $\langle I(x)\rangle$ for $a/d=0.2$,
$\delta/d=0.05$, and various values of the disordered segment length
$L/d=1200,1500,2250,3000$. (Recall that the incident mode impinges on the
disordered segment from the right end, $x=L$, which we have shifted to the
origin for the sake of clarity.) In all cases, the mean intensity decays
monotonically towards the exit of the disordered waveguide, the decay rate
being smaller the longer is the waveguide (provided that $L/\xi\gg 1$).
The contribution from resonances to the mean intensity, $\langle
I(x)\rangle_{reso}$, yielding transmission coefficients larger than $T_c=
0.4$, is shown in Fig.~\ref{fig_aix}(b). Broad distributions
are found with large maximum field intensities lying near the center of the
disordered waveguide. The contribution from typical realizations ($>90\%$),
$\langle I(x)\rangle_{typ}$ is plotted in Fig.~\ref{fig_aix}(c); a
qualitative behavior similar to that of the total mean intensity is
observed, except for a faster decay rate.

\begin{figure}[h]
\includegraphics[width=\columnwidth]{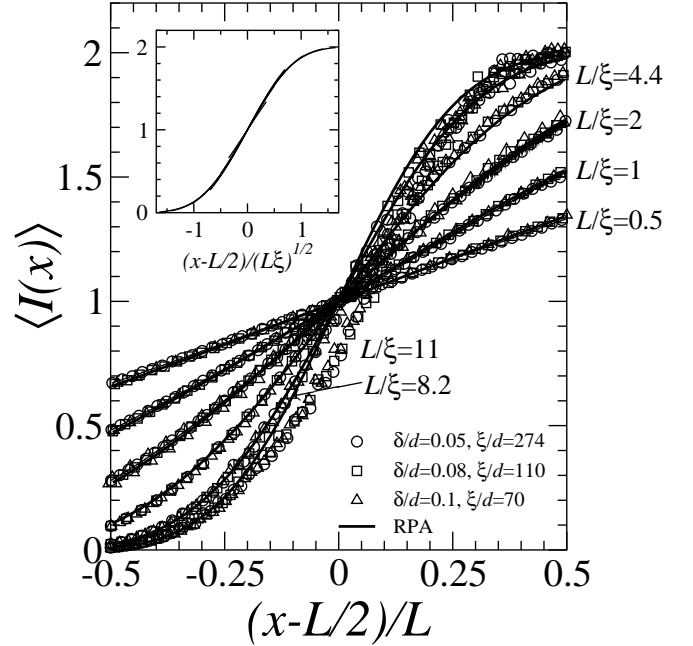}
\caption{Spatial distributions of average field intensity ($N=10^5$
realizations) as a function of renormalized position $\bar{x}\equiv(x-L/2)/L$
for $a/d=0.2$ and $\delta/d=$0.05 (circles), 0.08 (squares), and 0.1
(triangles). In each case, several disordered lengths are considered
according to $L/\xi=0.5,1,2,4.4,8.2,11$. Solid curves represent the
quasi-analytical, RPA results. Inset: the RPA results only as a function
of $\tilde{x}\equiv(x-L/2)/(L\xi)^{1/2}$.}
\label{fig_aix_col}
\end{figure}

In order to improve our understanding of the physics underlying the
formation of the field intensity patterns, we have replotted $\langle
I(x)\rangle$ by rescaling the $x$-dependence in units of the disordered
segment length $L$, $\bar{x}=(x-L/2)/L$. In addition to that, calculations
have been done for different roughness parameters $\delta/d=0.05,0.08$ and
0.1 (fixed $a/d=0.2$), by choosing $L$ in such a way that the ratio $L/\xi$
remains fixed (the corresponding values of the localization length $\xi$ are
given in Fig.~\ref{fig_alnt}). The resulting $\langle I(x)\rangle$ are
presented in Fig.~\ref{fig_aix_col}: The RPA quasi-analytical results
obtained from Eq.~(\ref{eq_ai_rpa}) are also included. 

Several conclusions can be drawn from the latter results. First,
$\langle I(\bar{x})\rangle$ exhibits in all cases a \textit{universal
behavior}, depending only on the ratio $L/\xi$ regardless of the
microscopic details of the 1D disorder.  Actually, as shown in the
inset in Fig.~\ref{fig_alnt}, we have observed that universality can
be pushed further, so that $\langle I(\tilde{x} )\rangle$ [with
$\tilde{x}=(x-L/2)/(L\xi)^{1/2}$] is a unique function. Second, for
moderate and even large $L/\xi$, the RPA expression predicts very
accurately the mean field distribution obtained numerically; a
monotonic decay from $\langle I(\bar{x}=0.5)\rangle=1+\langle
R(L)\rangle$ at the incoming end to $\langle
I(\bar{x}=-0.5)\rangle=\langle T(L)\rangle$, crossing the value
$\langle I\rangle=1$ through the middle of the disordered segment
$\bar{x}=0$, and being steeper the larger is $L/\xi$.
Third, deep into the 1D Anderson localization regime, $L/\xi\geq 11$
in Fig.~\ref{fig_aix_col}, the numerical results reveal a departure
from the RPA predictions, as evidenced by the shift of the $\langle
I\rangle=1$ crossing towards the incoming end.
We have investigated the physical origin of this discrepancy
by enforcing in the numerical calculations some of the assumptions
made in the RPA approach. First, uncorrelated disorder has been used
in the numerical calculations, with similar results to those for the
Gaussian correlation. Rapid phase averaging has also been carried out
at each realization prior to ensemble averaging, yielding no
significant differences. Thus neither finite correlation nor RPA can
give rise to the observed discrepancy.

\begin{figure}[h]
\includegraphics[width=\columnwidth]{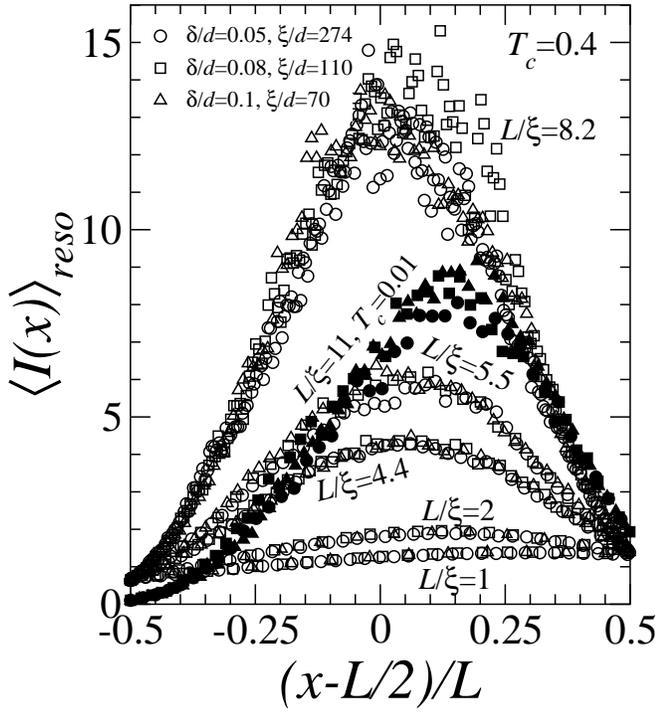}
\caption{Same as in Fig.~{\ref{fig_aix_col}} but for the contribution from
resonant realizations with $T_c=0.4$ (hollow symbols, $L/\xi
=1,2,4.4,5.5,8.2$) and $T_c=0.01$ (filled symbols, $L/\xi=11$). }
\label{fig_aix_reso_col}
\end{figure}

At this point, it is important to emphasize that plotted in Fig.
\ref{fig_aix_col} is the ensemble average of the intensity, which is a
non-self-averaging (strongly fluctuating) quantity. To gain insight
into the behavior of the field intensity pattern at different
individual realizations, we have separated typical and resonant
realizations according to a threshold value, $T_{c}$, of the mean
transmission coefficient. The contribution from resonances to the mean
intensity, $\langle I(x)\rangle _{reso}$, and (rescaled) $\langle
I(\bar{x} )\rangle _{reso}$, yielding transmission coefficients larger
than $T_{c}=0.4$, is shown in Figs.~\ref{fig_aix}(b) and
\ref{fig_aix_reso_col}.  One can see that the contribution from
resonances also exhibits universal behavior in the form of a broad
distribution with a relatively large maxima within the disordered
segment, being determined not only by the ratio $L/\xi$ (as in the
case of the total average), but also by the the cutoff parameter
$T_{c}$. Actually, from the comparison of the curves for $\langle
I(\bar{x} )\rangle _{reso}$ with different $T_{c}$ in
Fig.~\ref{fig_aix_reso_col}, it follows that $T_{c}$ fixes the
position of the maximum intensity, whereas the ratio $L/\xi$ sets the
precise value of the maxima. For fixed $T_{c}$, the maximum intensity
is higher for larger $L/\xi $; namely, stronger resonances are needed
for longer disorder in order to couple the same amount of energy
through the system (or similarly, to tunnel through a wider
barrier). Relaxing the definition of resonance (lowering $T_{c}$) for
fixed ratio $L/\xi$ (see Fig.~\ref{fig_aix_reso_col}), leads to
asymmetrical $\langle I(\bar{x})\rangle _{reso}$ distributions with
maxima shifted from the center to the incoming end of the disorder
segment.

\begin{figure}[h]
\includegraphics[width=\columnwidth]{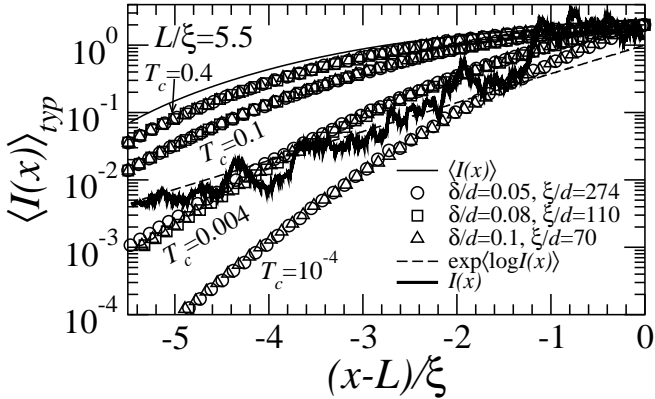}
\caption{Spatial distributions of the contribution to the average field 
intensity (log scale) from typical realizations with $a/d=0.2$,
$\delta/d=$0.05 (circles), 0.08 (squares), and 0.1 (triangles),
and fixed disordered length $L/\xi=5.5$,
for $T_c=10^{-4}$,  0.004, 0.1, 0.4, and 1 (the latter equivalent to 
$\langle I(x)\rangle$, thin solid curve). Also included are: 
$\exp\langle\log I(x) \rangle$ (dashed curve) and $I(x)$ (spatially
averaged in a log scale) for a single, 
typical realization (thick solid curve).}
\label{fig_aix_typ_col}
\end{figure}

The contribution to the average intensity from typical realizations
($>90\%$), $\langle I(x)\rangle _{typ}$, plotted in
Fig.~\ref{fig_aix}(c), also depends universally on $L/\xi $ and
$T_{c}$ (not shown here), and shows a qualitative behavior similar to
$\langle I(x)\rangle $, with a faster decay, as
expected. Interestingly, neither $\langle I(x)\rangle $ nor $\langle
I(x)\rangle _{typ}$ decay exponentially, but rather manifest a
$s$-like dependence, as mentioned above. This means that, no matter
how long the realization is [i.e., how small is $\exp (-L/\xi )$], a
lengthening of the system (increase of $L$ with the localization
length kept fixed) leads, surprisingly enough, to essential changes in
the energy distribution inside. Namely, provided that the strength of
the disorder is fixed, the longer a randomly disordered sample is, the
slower is the decay of the intensity [both $\langle I(x)\rangle $ and
$\langle I(x)\rangle _{typ}$] from the incoming end deep into the
sample [this is neatly observed in Figs.~\ref{fig_aix}(a) and
~\ref{fig_aix}(c)].  In other words, the ''penetration depth'' of both
$\langle I\rangle$ and $\langle I\rangle _{typ}$ into a 1D random
system is independent of the strength of scattering. This effect is,
however, dependent on the value $T_{c}$ in the definition of $\langle
I\rangle _{typ}$, as illustrated in Fig.~\ref{fig_aix_typ_col}: with
the cutoff decreasing, the slowly decaying part of $\langle
I(x)\rangle _{typ}$ near the incoming end ($x=L$) diminishes, the
distribution thus decaying more abruptly. Obviously, the longer a
realization is, the smaller $T_{c}$ is necessary for the transition to
take place. Interestingly, the intensity for a single typical
realization for which $T\sim \langle T\rangle$ appears to
decay approximately exponentially $I\sim \exp (-x/\xi )$, as seen in
Fig.~\ref{fig_aix_typ_col} (its oscillations are smoothed spatially on
a log scale).  This is in accordance with the behavior of the average
logarithm of the intensity, which fluctuates less strongly than the
intensity itself and fits very accurately $\langle\log I(x)\rangle
\simeq -|x-L|/\xi$ (see Fig.~\ref{fig_aix_typ_col}), revealing its
self-averaging nature.

\begin{figure}[h]
\includegraphics[width=\columnwidth]{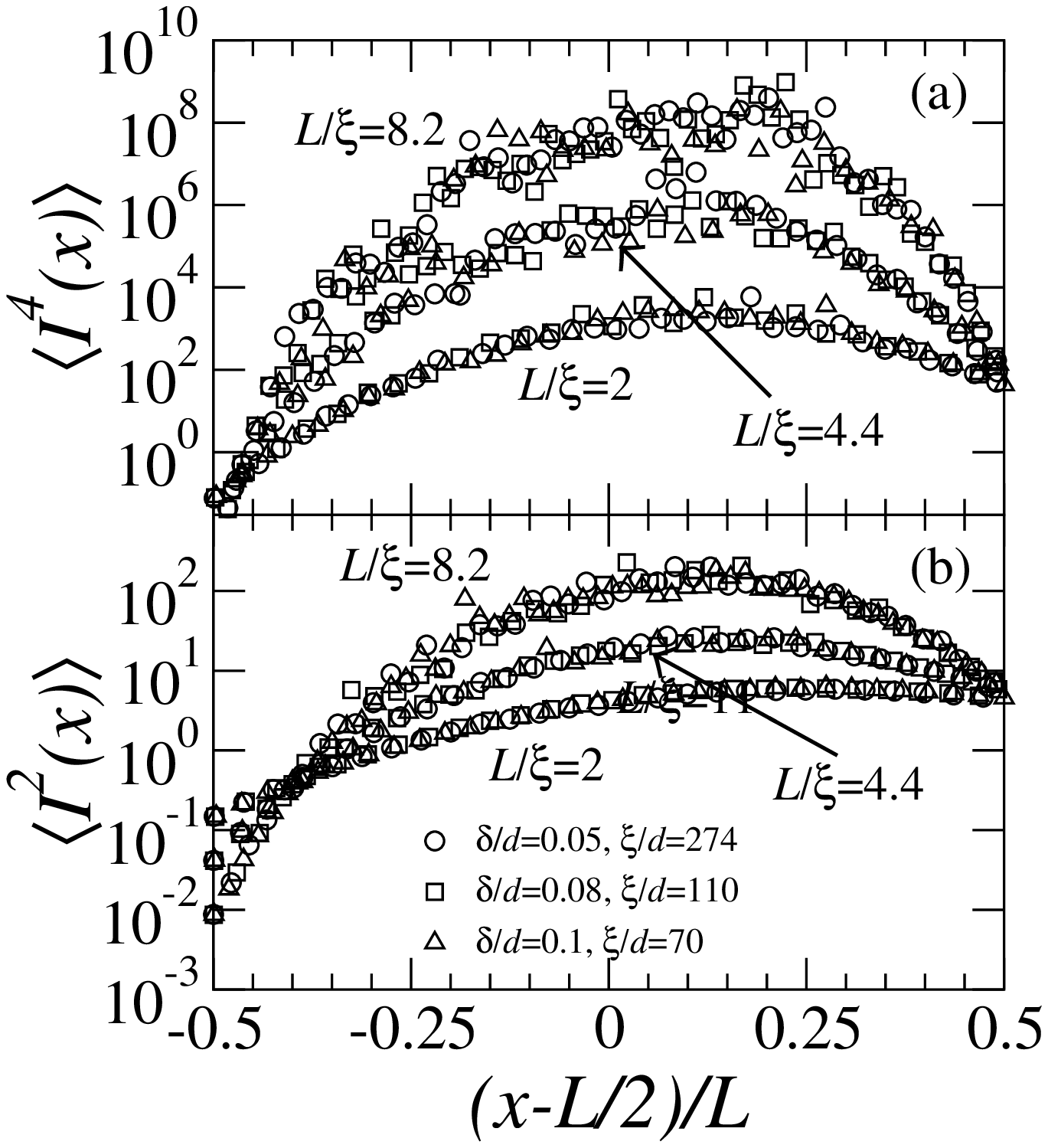}
\caption{Same as in Fig.~{\ref{fig_aix_col}} but for the 4th (a) and 2nd 
(b) moments of the field intensity, and $L/\xi=2,4.4,8.2$. }
\label{fig_aix_n}
\end{figure}

Finally, we have calculated higher-order moments of the mean intensity
$\langle I^n(\bar{x})\rangle$. In Fig.~\ref{fig_aix_n}, the numerical
results are shown in the case $n=2,4$ for some of the disordered
waveguides considered above. The most remarkable feature is the broad,
resonant-like shape, revealing the increasing (for higher $n$)
influence of (low-probability) resonances, with huge field
intensities.

\section{CONCLUDING REMARKS}

\label{sec_con}

To summarize, we have developed a formalism to calculate the field inside
surface-disordered waveguides, similar to that of the invariant embedding
equations for the reflection and transmission coefficients. By applying it
to 2D single-mode waveguides with planar walls and Gaussian-correlated
surface roughness, we have investigated the occurrence of resonances in the
1D Anderson localization regime, with emphasis on the resulting field
intensity distribution both for given realizations and ensemble averages.

We have examined the frequency dependence %(in a narrow frequency range)
of the transmission coefficient $T(\omega)$ for different realizations; 
it exhibits well-defined resonance-type behavior inherent to the
localization regime. This enables us to separate typical realizations,
characterized by very low (as expected from the average $\langle \log
T\rangle\sim -L/\xi $) values of $T$ and a monotonically decaying
intensity, from resonances with transmission coefficients close to one
and extremely high intensity maxima (localization) in a region around
the center of the system.

Numerical simulation calculations for the mean field intensity
$\langle I(x)\rangle$ along the disordered segment of the waveguide
reveal a universal behavior completely determined by the ratio
$L/\xi$: A smooth decay from the initial value of $\langle I
\rangle\sim 1+\langle R(L)\rangle$ at the incoming end, to the
outgoing mean transmitted field intensity $\langle I\rangle\sim
\langle T(L)\rangle$, crossing the value $\langle I\rangle\sim 1$
at/near the center of the disordered segment. For moderately strong
disorder $L/\xi\overset{>}{\sim} 1$, the quasi-analytical (RPA)
prediction (\ref{eq_ai_rpa}) fully agrees with the numerical
calculations.  However, for strong disorder $L/\xi\gg 1$, the
numerical results exhibit, unlike the RPA result, a shift of the
mid-point ($\langle I\rangle\sim 1$) towards the incoming edge.
The contribution to $\langle I(x)\rangle$ from resonant realizations
(those yielding anomalously large transmission above a threshold value
$T_c$) manifests also universality characterized by the parameters
$L/\xi$ and $T_c$: Its shape is a broad distribution whose maximum
value, which is larger for stronger disorder, shifts from the center
towards the incoming edge with decreasing $T_c$. On the other hand, we
have found that the contribution from such low-probability resonances
become more dramatic in higher-order moments of the total intensity
distribution.

The contribution from typical realizations to the total average,
$\langle I(x)\rangle_{typ}$, depends on the cutoff value $T_{c}$. For
$T_{c}$ not too small, $\langle I(x)\rangle_{typ}$ (as well as
$\langle I(x)\rangle$) inside a 1D random system is slightly dependent
on the strength of the scattering, and increases with the increase of
the total length, $L$, of the system. With decreasing threshold value
$T_{c}$, the penetration depth ceases to depend on $L$ and $\langle 
I(x)\rangle _{typ}$ decays more rapidly. In this regard, 
evidence of the self-averaging nature of $\log I(x)/x$  is given by the 
behavior of $\log I(x)$ for single, typical realizations, and also by
the result that $\langle\log I(x)\rangle\simeq -|x-L|/\xi$.

\begin{acknowledgments}
This work was supported in part by the Spanish Direcci{\'o}n General de
Investigaci{\'o}n (Grants BFM2000-0806 and BFM2001-2265), and by
the ONR Grant ONR\#N000140010672.
\end{acknowledgments}

%%%%%%%%%%%%%%%%%%%%%%%%%%%%%%%%%%%%%%%%%%%

\begin{thebibliography}{99}

\bibitem{frisch} U. Frisch, C. Froeschle, J.-P. Scheidecker, and P.-L. Sulem,
 \pra{\bf 8}, 1416 (1973).
\bibitem{papanico} G. Papanicolaou, J. Appl. Math. {\bf 21}, 13 (1971); W.
Kohler, G. Papanicolaou, J. Math. Phys. {\bf 14}, 1753 (1973); W. Kohler, G.
Papanicolaou, J. Math. Phys. {\bf 15}, 2186 (1974); J. Keller, G.
Papanicolaou, J. Weilenmann, J. Appl. Math. {\bf 32}, 583 (1978).
\bibitem{Azbel} M. Ya. Azbel, \prb{\bf 22}, 4045 (1980); M. Ya.
Azbel, \prl{\bf 47}, 1015 (1981); M. Ya. Azbel, P. Soven, \prl{\bf 49}
, 751 (1982);  M. Ya. Azbel, \prb{\bf 28}, 4106 (1983);  M. Ya. Azbel, P.
Soven, \prb{\bf 27}, 831 (1983); M. Ya. Azbel, \prb{\bf 27, }3901
(1983); E. Cota, J. V. Jose, and M. Ya. Azbel, \prb{\bf 32}, 6157 (1985)
\bibitem{klbook} V. Kliatzkin, \textit{Stochastic Equations and Waves in
Random Media} (Nauka, Moscow, 1980), \textit{in Russian}.
\bibitem{lifsh} I. M. Lifshits, S. A. Gredeskul, and L. A. Pastur, 
\textit{Introduction to the Theory of Disordered Systems} 
(Wiley, New York, 1988), Chap. 7.
\bibitem{sheng} \textit{Scattering and Localization of Classical Waves in
Random Media}, edited by P. Sheng (World Scientific, Singapore, 1990).
\bibitem{frei} V. Freilikher and S. Gredeskul, Progress in Optics 
               \textbf{30}, 137 (1992).
\bibitem{letho} V. S. Letokhov, Sov. Phys. JETP 26, 835 (1968).
\bibitem{cao} H. Cao, Y. G. Zhao, S. T. Ho, E. W. Seelig, Q. H. Wang, and
              R. P. H. Chang, \prl{\bf 82}, 2278 (1999); 
              H. Cao, Y. Ling, J. Y. Xu, C. Q. Cao, and P. Kumar, 
              \prl{\bf 86}, 4524 (2001).
\bibitem{natu} D. S. Wiersma, Nature \textbf{406}, 132 (2000); 
               D. S. Wiersma, S. Cavalieri, Nature \textbf{414}, 708 (2001).
\bibitem{lag01} G. van Soest, F. J. Poelwijk, R. Sprik, and Ad Lagendijk,
                 \prl{\bf 86}, 1522 (2001).
\bibitem{prl98} J. A. S\'anchez-Gil, V. Freilikher, I. V. Yurkevich, 
                and A. A. Maradudin, \prl{\bf 80}, 948 (1998).
\bibitem{mnvprl98} A. Garc{\'{\i}}a-Mart{\'{\i}}n, J. A. Torres, 
 J. J. S\'aenz, and M. Nieto-Vesperinas, \prl{\bf 80}, 4165 (1998); 
 A. Garc{\'{\i}}a-Mart{\'{\i}}n, T. L\'opez-Ciudad, J. J. S\'aenz, 
and M. Nieto-Vesperinas, \prl{\bf 81}, 329 (1998).
\bibitem{prb99} J. A. S\'anchez-Gil, V. Freilikher, A. A. Maradudin,
and I. Yurkevich, \prb{\bf 59}, 5915 (1999).
\bibitem{mnvprl00} A. Garc{\'{\i}}a-Mart{\'{\i}}n, J. J. S\'aenz, and M.
                   M. Nieto-Vesperinas, \prl{\bf 84}, 3578 (2000).
\bibitem{mnvprl01} A. Garc{\'{\i}}a-Mart{\'{\i}}n and J. J. S\'aenz,
                   \prl{\bf 87}, 116603 (2001).
\bibitem{mnvprl02} A. Garc{\'{\i}}a-Mart{\'{\i}}n, F. Scheffold, 
                   M. Nieto-Vesperinas, and J. J. S\'aenz,
                   \prl{\bf 88}, 143901 (2002).
\bibitem{izma02} F. M. Izrailev and N. M. Makarov, \ol{\bf 26}, 1604 (2002).
\bibitem{emb} N. Makarov and I. Yurkevich, Zh. {\'{E}}ksp. Teor. Fiz. 
\textbf{96}, 1106 (1989) [Sov. Phys. JETP \textbf{69}, 628 (1989)]; A.
Krokhin, N. Makarov, V. Yampolskii, and I. Yurkevich, Physica B 
\textbf{165\&166}, 855 (1990); V. Freilikher, M. Pustilnik, and I.
Yurkevich, \prl{\bf 73}, 810 (1994).

\end{thebibliography}
\end{document}